\documentclass{article}
\usepackage[T1]{fontenc} % add special characters (e.g., umlaute)
\usepackage[utf8]{inputenc} % set utf-8 as default input encoding
\usepackage{ismir,amsmath,cite,url}
\usepackage{graphicx}
\usepackage{color}

\usepackage{times}
\usepackage{latexsym}
\usepackage{bm}
\usepackage{tabularx,booktabs}
\usepackage{multirow}
\usepackage{inconsolata}
\usepackage{caption}
\usepackage{subcaption}
\usepackage{makecell}
\usepackage{lipsum}
\usepackage[hang,flushmargin]{footmisc}
\usepackage{microtype}
\usepackage{lineno}

\title{Practical and Reproducible Symbolic Music Generation by Large Language Models with Structural Embeddings}

\multauthor
{Seungyeon Rhyu$^{1,2,\dagger,\star}$\thanks{$\dagger$ Work
completed during an internship at LG AI Research and Ph.D. candidate at Seoul National University.}\thanks{$\star$ Seungyeon Rhyu and Jaehyeon Kim are currently working at Pozalabs and Krafton, respectively. For inquiries, please contact Seungyeon Rhyu at <rsy1026@snu.ac.kr>.}  \hspace{1cm} Kichang Yang$^1$ \hspace{1cm} Sungjun Cho$^1$} { \bfseries{Jaehyeon Kim$^{1,\star}$ \hspace{1cm} Kyogu Lee$^{2,3}$ \hspace{1cm} Moontae Lee$^1$}\\
$^1$ LG AI Research\\
$^2$ Music and Audio Research Group (MARG), Seoul National University, South Korea\\
$^3$ Graduate School of AI, AI Institute, Seoul National University, South Korea
}

\def\authorname{S. Rhyu, K. Yang, S. Cho, J. Kim, K. Lee and M. Lee}
\usepackage[bookmarks=false,pdfauthor={\authorname},pdfsubject={\papersubject},hidelinks]{hyperref}

\begin{document}

\maketitle
\begin{abstract}
Music generation introduces challenging complexities to large language models. Symbolic structures of music often include vertical harmonization as well as horizontal counterpoint, urging various adaptations and enhancements for large-scale Transformers. However, existing works share three major drawbacks: 1) their tokenization requires domain-specific annotations, such as bars and beats, that are typically missing in raw MIDI data; 2) the pure impact of enhancing token embedding methods is hardly examined without domain-specific annotations; and 3) existing works to overcome the aforementioned drawbacks, such as MuseNet, lack reproducibility. To tackle such limitations, we develop a MIDI-based music generation framework inspired by MuseNet, empirically studying two structural embeddings that do not rely on domain-specific annotations. We provide various metrics and insights that can guide suitable encoding to deploy. We also verify that multiple embedding configurations can selectively boost certain musical aspects. By providing open-source implementations via HuggingFace\footnote{\url{https://github.com/anonymous-submission-351532/examuse-transformers}}, our findings shed light on leveraging large language models toward practical and reproducible music generation.
\end{abstract}

\section{Introduction}\label{sec:introduction}
Symbolic music generation is a rapidly growing field in machine learning, as reflected by the continuously increasing number of musical datasets that provide a solid foundation for training large music generation models\cite{gemmeke2017audio,agostinelli2023musiclm,simon2019pianotransformer}. While the sequential aspect of music naturally led to the application of the Transformer architecture\cite{vaswani2017attention} to the musical domain, the vertical (e.g., multiple notes can be played simultaneously) and temporal (e.g., every note played has a certain duration) structures in musical data call for a careful tokenization procedure that is distinct from the typical positional encoding used in text generation\cite{wang2021musebert}.

\begin{figure}[!t]
\begin{subfigure}{=\columnwidth}
  \centering
  \includegraphics[width=.9\columnwidth]{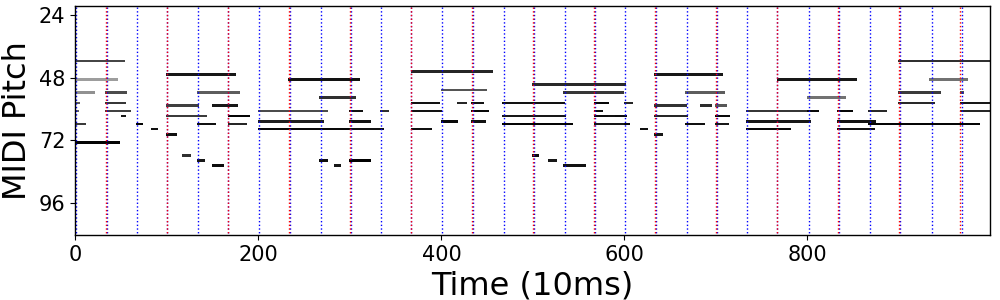}
  \vspace{-.05in}
  \caption{POP909}
  \label{fig:beat_good}
\end{subfigure}
\begin{subfigure}{=\columnwidth}
  \centering
  \includegraphics[width=.9\columnwidth]{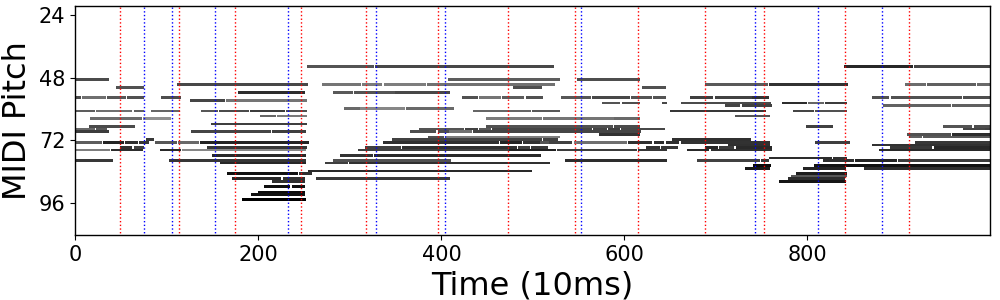}
  \vspace{-.05in}
  \caption{GiantMIDI}
  \label{fig:beat_bad}
\end{subfigure}%
\caption{Example of beat detection results using the famous Madmom algorithm\cite{bock2016madmom} on (a) POP909 and (b) GiantMIDI. The blue and red dotted lines indicate ground truth and predicted beats, respectively.}
\label{fig:beat}
\vspace{-.1in}
\end{figure}

\begin{figure*}
    \centering
    \includegraphics[width=.9\linewidth]{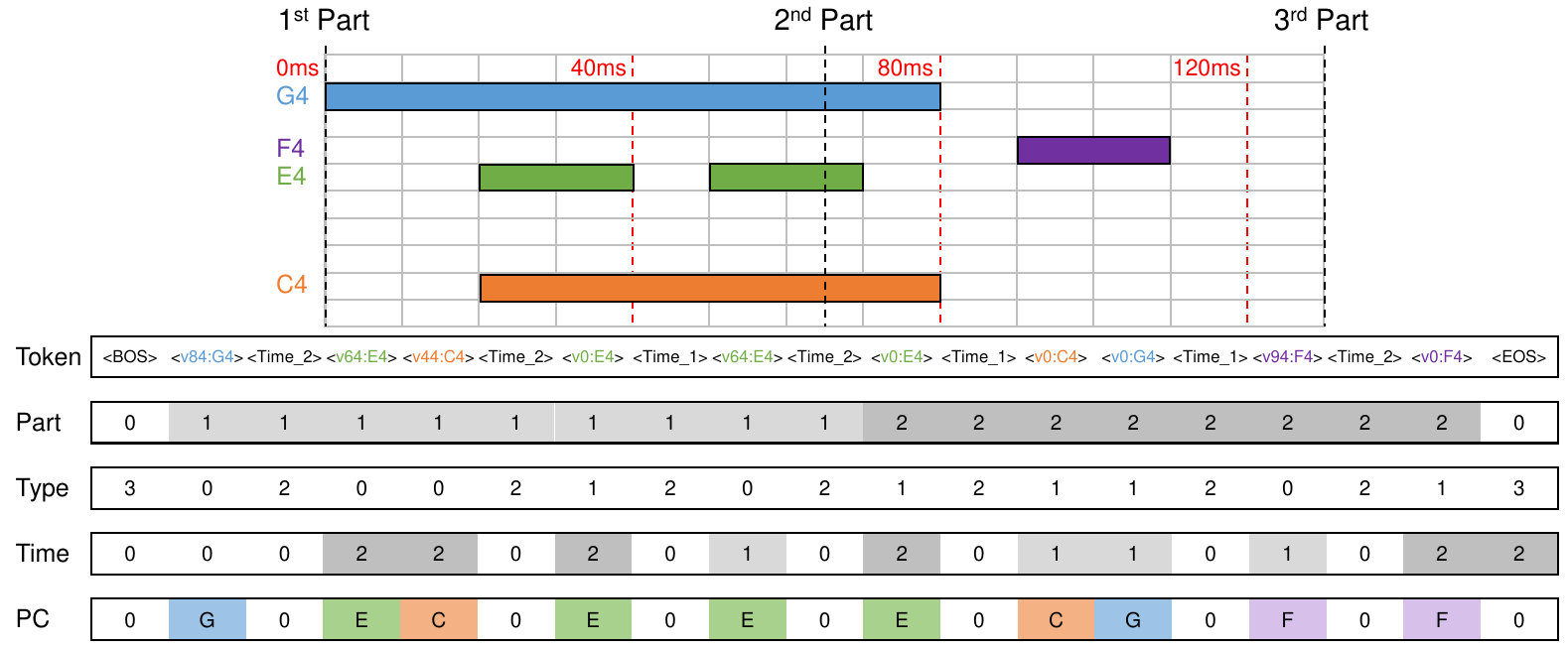}
    \caption{Illustration of the reconstructed structural embeddings. \textbf{Top:} An example of a piano roll where 5 notes appear in different pitches and timings. \textbf{Bottom:} The corresponding tokens and structural embeddings.}
    \label{fig:embeddings}
\vspace{-.1in}
\end{figure*}

While there exist various tokenization techniques such as REMI\cite{huang2020pop} and CP\cite{hsiao2021compound}, the MIDI-like tokenization\cite{huang2018music} is considered most compatible with large-scale musical data from cloud-based platforms (e.g. MIDI transcription from YouTube audio). This MIDI data may not include detailed annotations of the temporal division of bars.
While these annotations can be generated manually\cite{wu2020jazz, foscarin2020ASAP}, this process comes at a high cost. Another approach is to use heuristic algorithms\cite{huang2020pop, zhang2022structure}, but these methods can only be applied reliably when the input music progresses at a constant speed or tempo, with well-separated bars and beats\cite{wang2020pop909}. Unfortunately, this is not the case for the majority of MIDI data directly transcribed from live music audio (see \figref{fig:beat}), due to performers frequently adding note on- or off-sets that deviate from the scripted tempo\cite{kim2013statistical, grosche2010what}.

Nonetheless, enhancing the vanilla MIDI tokenization to effectively encode musical structure without domain-specific annotations is not yet studied in depth. One exemplar work is MuseNet, which introduced additional \textit{structural} embeddings to better derive the structural context of music from MIDI-based tokens, demonstrating state-of-the-art performance in prompt-based music generation\cite{payne2019musenet}. However, the reasoning behind its design choice of structural embeddings is still unclear. There is no clear experimentation to validate their effects separately from the overall capacity of the model architecture. Furthermore, MuseNet has not yet published any open-source implementations to support further tuning and testing. As a result, it is uncertain whether incorporating these supplementary embeddings is truly practical and optimal in capturing structural contexts of raw MIDI music, which hinders potential applications with data that lacks domain annotations.

In response to such limitations, we focus on the structural embeddings proposed by MuseNet and deeply investigate their effect on encoding musical structures based on raw MIDI data. Specifically, we train a vanilla GPT-2 model\cite{radford2019language} equipped with our implementation of the structural embeddings from MuseNet. 
We build these structural embeddings upon practical methods considering the noisiness of raw MIDI data. We also perform ablation studies by adjusting different initialization methods to study their impact on generation performance. With this paper, we aim to provide practitioners with insights into how to effectively encode structural information in music without manual annotations, as well as an accessible pipeline to train models on large music datasets.
Our main contributions are as follows:
1) We focus on the effect of structural embeddings for learning musical structure from the raw MIDI data without domain-specific annotations;
2) We implement the supplementary structural embeddings suggested by MuseNet using practical methods that fit the raw MIDI data. For public use, we open-source our code and a demo page;
3) We empirically analyze and clearly validate the impact of different initializations of structural embeddings via substantial evaluation.

\section{Method}\label{sec:method}
\begin{figure*}[!t]
    \centering
    \includegraphics[width=\textwidth,trim={2.4cm 3.5cm 14.1cm 2.4cm},clip]{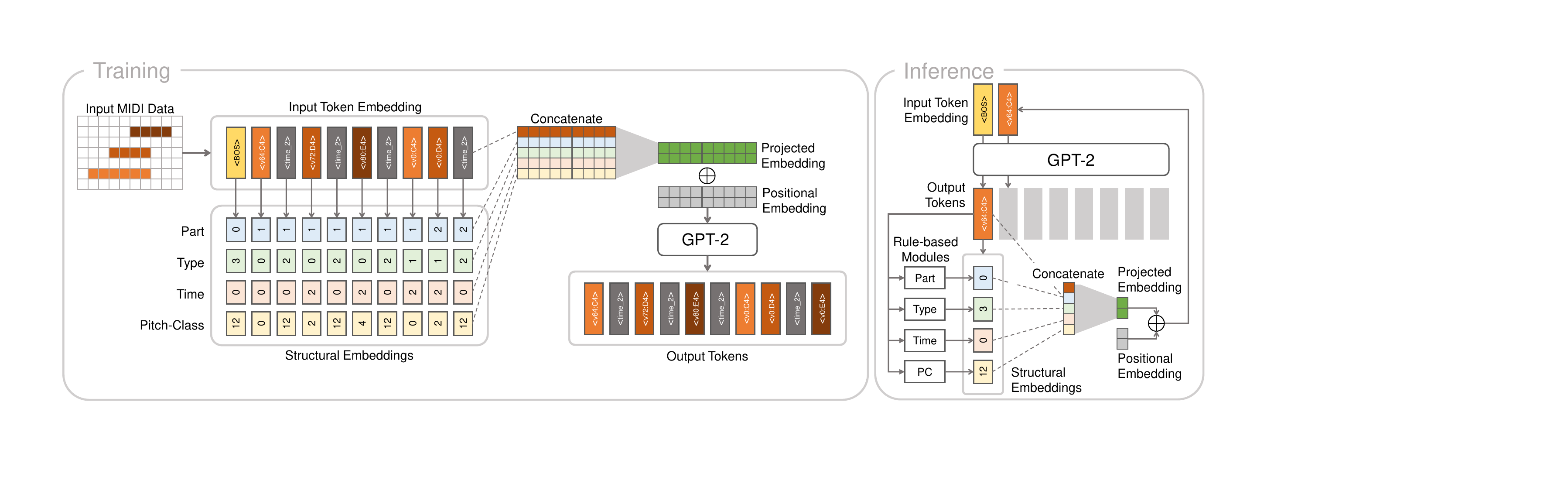}
    \vspace{-.25in}
    \caption{Illustration of our music generation framework. \textbf{Left:} During training, four types of structural embeddings are concatenated to the input tokens, then projected to the input embedding size for next-token prediction. \textbf{Right:} During inference, structural information inferred via rule-based modules is used for autoregressive generation.}
    \label{fig:method}
\vspace{-.1in}
\end{figure*}

To encode musical data as a sequence of tokens, we use the event-based tokenization method of MuseNet\cite{payne2019musenet, huang2018music}. We discard any meta-information and use only 4,196 different input tokens: 4,096 note-on tokens resulting from 32 quantized velocities and 128 pitch numbers (e.g. \texttt{<v64:C4>}), 128 note-off tokens with zero velocity (e.g. \texttt{<v0:C4>}), and 100 time-shift tokens representing 100 divisions from 10 ms to 1000 ms (e.g. \texttt{<wait:2>}). We also use special \texttt{<BOS>}, \texttt{<EOS>}, and \texttt{<PAD>} tokens as we need to batch pieces with varying durations into a fixed length.

\subsection{Structural Embeddings}\label{sec:structural_embeddings}

We revisit the original structural embeddings of MuseNet \cite{payne2019musenet} and reconstruct them to capture meaningful structures in raw MIDI data. We implement the four structural embedding layers, which are named as part, type, time, and pitch-class (PC), modifying the MuseNet version towards better fitness to audio-based MIDI data. Among the four embeddings, the part embedding is the only one that is identically reconstructed from MuseNet, while others are modified or re-interpreted in our application. As demonstrated in \figref{fig:embeddings}, four additional embedding layers construct the structural embeddings which are concatenated with the token embeddings, projected to the hidden size of the model, and added to the positional embedding. We expect from this pipeline that each structural embedding on each token is considered equally important with the token embedding. Below we provide a detailed description of each type of structural embedding:

\noindent\textbf{Part.} 
The part embedding informs which temporal part among the entire song each token is located at. We divide the total time length of a song equally by 128. For each event token, we compute a part class as one of the 128 temporal parts of the sequence. As a result, each token is indexed starting from \texttt{0} to \texttt{128}, where \texttt{0} indicates the non-note tokens (e.g. \texttt{<BOS>}, \texttt{<EOS>}) and \texttt{1} indicates the actual starting part of the song. 

\noindent\textbf{Type.} 
The type embedding indicates the type of each token, whether the token is related to note-on(0), note-off(1), time(2), and other types(3) of event. This can help the model distinguish different token types more easily and learn dependencies among tokens within the same token type. Note that without the type embedding, all tokens are learned independently of each other regardless of their inherent type. 

\noindent\textbf{Time.} 
The time embedding informs the model time difference in 10ms between each non-time-shift token and its previous non-time-shift token. It is similar to Musenet, yet we re-interpreted the original to represent the \textit{relative} time between the tokens. We directly allocate this information to the note-on or -off tokens which follow each time-shift token to explicitly encode the temporal concurrency or progression of notes. If multiple time-shift tokens are interjected, we only count the closest time-shift token. It is to bound the class size as the time differences may be unbounded. We expect the original event tokens to give extra information. Once a time token appears in a sequence, the corresponding time-shift value (i.e. 2 if \texttt{<time\_2>}) is assigned as a time class to the following non-time-shift tokens. The time-shift tokens are assigned 0.

\noindent\textbf{Pitch-Class (PC).} 
The pitch-class embedding informs the pitch-class of each note-on or -off token. At every note-on and note-off token, the corresponding pitch-class is directly applied as a PC class, while other tokens are assigned 0. MuseNet originally proposed to use the relative position of each note within the corresponding chord \cite{payne2019musenet}. 
However, it is tricky to determine the boundary of a chord due to the notes sequentially intermeshed or sustained by the piano pedals as shown in \figref{fig:beat_bad}. Therefore, we instead utilize the pitch-class of each note-related token that can explicitly represent the tonal color of each token. We conjecture that incorporating pitch-class embedding allows the model to stack tonal information via self-attention and thus immediately capture the harmonic context of the input.

\subsection{Training and Inference}

As our backbone architecture, we use GPT-2\cite{radford2019language} following MuseNet\cite{payne2019musenet}. However, we do not utilize sparse attention\cite{child2019generating} and mixup\cite{zhang2017mixup} to focus on the impact of adjusting structural embeddings for musical data without any trade-offs in computational efficiency or regularization. The overall procedure for training and inference using our model is illustrated in \figref{fig:method}. We concatenate the four structural embeddings with input token embeddings along the feature axis and project the resulting embeddings into the model’s hidden dimension using a fully-connected layer. We also add a trainable, randomly initialized sequence positional encoding to the resulting embeddings.

For training, we use a next-token prediction task, analogous to language modeling\cite{radford2019language}. Let $X=[x_1, …, x_N]$ represent a sequence of input tokens, where $N$ denotes the sequence's length. At each step, the model learns to predict $x_i$ from $x_{<i}$ by optimizing $\mathcal{L} = -\sum{\log{p(x_n|x_{<i})}}$. During inference, the model autoregressively generates the sequence given a prompt of MIDI tokens. After generating each token, we extract its four structural labels—part, type, time, and PC—from all subsequent tokens using rule-based modules. These labels are then fed into the model for the next step in generation.

\subsection{Initialization Methods}\label{sec:initialized_methods}

In this paper, we examine the effect of two different initialization methods of structural embeddings. The first method is to use truncated normal initialization\cite{wolf2019huggingface}. The second method involves sinusoidal initialization, applied only to the time-related embeddings. This approach, inspired by Guo \textit{et al}\cite{guo2022domain}, employs sinusoidal encoding to preserve the continuity of ordinal musical attributes, such as note onset and pitch. We similarly use sinusoidal lookup tables for temporal embeddings, part and time, as follows: 
\begin{equation}
\begin{aligned}
    SE_{(k, 2i)} &= \sin(k/(10000/w)^{2i/d}) \\
    SE_{(k, 2i+1)} &= \cos(k/(10000/w)^{2i/d})
\end{aligned}
\end{equation}
where $k$ is a class index for part or time, $i$ is a feature index, $d$ is the hidden size, and $w$ is a scaling factor. By using two different values for $w$, we ensure that the corresponding embeddings are orthogonal, allowing the part and time attributes to be represented without interference\cite{guo2022domain}. Considering that Part is composed of a much larger unit than Time which is in 10ms, we manually set $w=10$ and $w=1$, respectively. We also note that the average length of each part is around 432 ms (the average song length is $55.39\pm 25.23$ seconds). By testing different initializations, we expect to examine whether taking temporal continuity into account affects generative performance under the nature of musical attributes.

\begin{table*}
    \centering
    \resizebox{.95\linewidth}{!}{
    \begin{tabularx}{\linewidth}{c*{11}{X}}
        \toprule
        \multicolumn{1}{l|}{Metric} & \multicolumn{1}{l|}{Prompt} & $\mathcal{SI}_{3}^{8}$ & $\mathcal{SI}_{8}^{15}$ & $\mathcal{SI}_{15}^{N}$ & $\mathcal{CPVR}_{2}$ & $\mathcal{CPVR}_{3}$ & $\mathcal{CPVR}_{4}$ & $\mathcal{CPI}_{2}$ & $\mathcal{CPI}_{3}$ & $\mathcal{CPI}_{4}$\\ 
        \midrule
        \multicolumn{1}{l|}{GPT2} & \multicolumn{1}{l|}{16} & 0.2025  &  0.1801  & 0.1653 & 0.4628     &0.5011   &  0.4766 & 0.3910  &  0.5367  &  0.6331 \\
        \multicolumn{1}{l|}{GPT2-RE} & \multicolumn{1}{l|}{16} & 0.1908  &  0.1707 &  0.1585 &  0.4525  &   0.4880  &   0.4630 & \textbf{0.4064}  &  \textbf{0.5578} &   \textbf{0.6570} \\
        \multicolumn{1}{l|}{GPT2-SE} & \multicolumn{1}{l|}{16} & \textbf{0.2169}  &  \textbf{0.1948} &  \textbf{0.1816} & \textbf{0.5216}   &  \textbf{0.5682}  &   \textbf{0.5460} & 0.3162  &  0.4389  &  0.5222 \\
        \midrule
        \multicolumn{1}{l|}{GPT2} & \multicolumn{1}{l|}{64} & 0.1985  &  0.1767  & 0.1613 & 0.4576     &0.4960   &  0.4703 & \textbf{0.4034}  &  \textbf{0.5500}  &  \textbf{0.6463} \\
        \multicolumn{1}{l|}{GPT2-RE} & \multicolumn{1}{l|}{64} & 0.2036  &  0.1832 &  0.1691 & 0.4915   &  0.5318  &   0.5087 & 0.3672  &  0.5057  &  0.5979 \\
        \multicolumn{1}{l|}{GPT2-SE} & \multicolumn{1}{l|}{64} & \textbf{0.2090}  &  \textbf{0.1905} &  \textbf{0.1774} & \textbf{0.5092}   &  \textbf{0.5540}  &   \textbf{0.5305} & 0.3372  &  0.4644 &   0.5502 \\
        \midrule
        \multicolumn{1}{l|}{GPT2} & \multicolumn{1}{l|}{256} & 0.1874  &  0.1689  & 0.1555 & 0.4460     &0.4808   &  0.4544 & \textbf{0.4320}  &  \textbf{0.5844}  &  \textbf{0.6834} \\
        \multicolumn{1}{l|}{GPT2-RE} & \multicolumn{1}{l|}{256} & 0.1941  &  0.1763 &  0.1625 & 0.4755    & 0.5126   &  0.4882 & 0.3977  &  0.5426 &   0.6373 \\
        \multicolumn{1}{l|}{GPT2-SE} & \multicolumn{1}{l|}{256} & \textbf{0.2015} &   \textbf{0.1816}  & \textbf{0.1686} & \textbf{0.4905}   &  \textbf{0.5321} &   \textbf{0.5079} & 0.3708  &  0.5064  &  0.5960 \\
        \bottomrule
    \end{tabularx}
    }
    \caption{Results of the objective evaluation scores on the three metrics: Structureness Indicators (SI), Chord Progression Variation Rationality (CPVR), and Chord Progression Irregularity (CPI).}
    \label{tab:objective}
\end{table*}%

\begin{figure}
\begin{subfigure}{.33\columnwidth}
  \centering
  \includegraphics[width=.99\columnwidth]{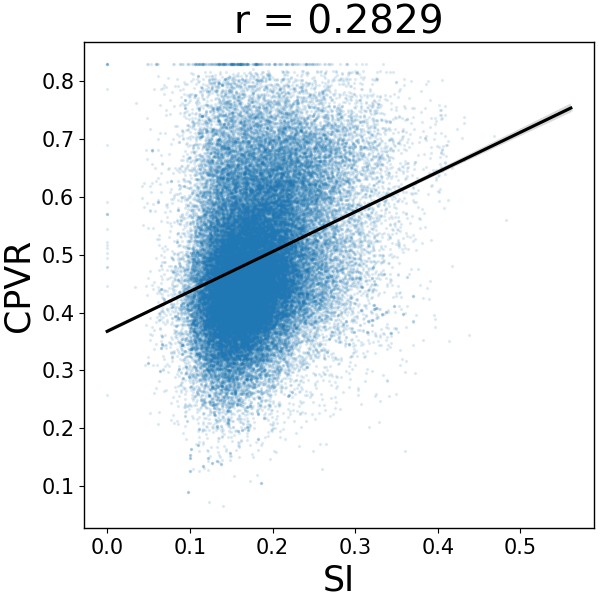}
  \caption{SI vs. CPVR}
  \label{fig:si-cpvr}
\end{subfigure}%
\begin{subfigure}{.33\columnwidth}
  \centering
  \includegraphics[width=.99\columnwidth]{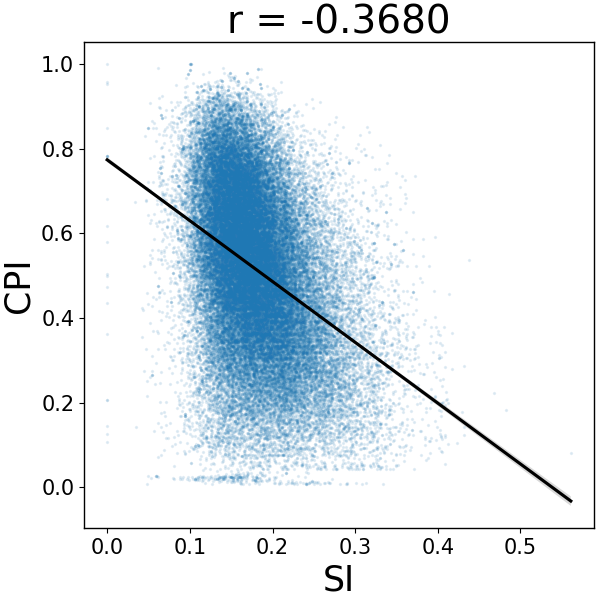}
  \caption{SI vs. CPI}
  \label{fig:si-cpi}
\end{subfigure}%
\begin{subfigure}{.33\columnwidth}
  \centering
  \includegraphics[width=.99\columnwidth]{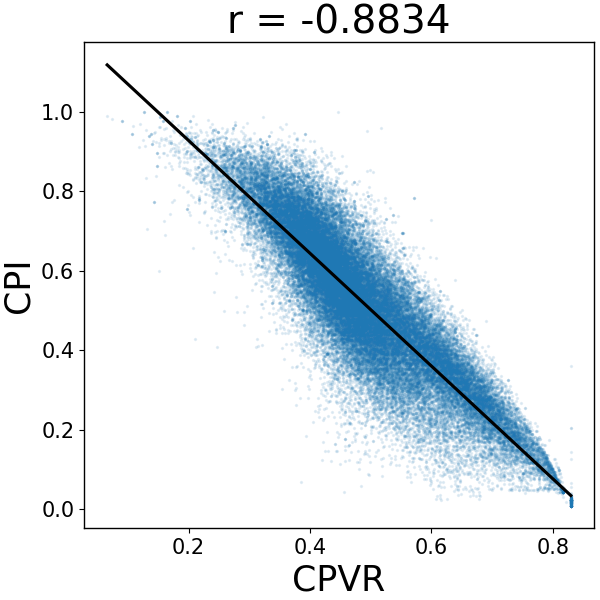}
  \caption{CPVR vs. CPI}
  \label{fig:cpvr-cpi}
\end{subfigure}%
%\caption{Evaluation results for four musical metrics: SI, CPR($n=2$), CPR($n=3$), CPR($n=4$)}
\caption{Correlation results among SI, CPVR, and CPI computed from the three metric scores of all models. Pearson's correlation coefficients are also reported on the top of each plot.}
\label{fig:corr}
\vspace{-.1in}
\end{figure}

\section{Experiments}\label{sec:experiments}
\subsection{Experimental Setup}\label{sec:experimental_setup}

\noindent\textbf{Datasets.} 
For training, we use two datasets comprising 1,748 pieces from Pop1k7\cite{hsiao2021compound} and 5,842 pieces from GiantMIDI-Piano\cite{kong2020giantmidi}. Both datasets consist of transcribed MIDI files from large-scale piano performances. For computational efficiency, we exclude pieces from GiantMIDI-Piano that are longer than 10 minutes. We also apply 7 pitch transpositions up to $\pm 3$ half steps and 5 time stretches up to $\pm 0.05$ \cite{huang2018music}, consequently using a total of 17,383 hours of transcribed MIDI data. A small portion ($0.01\%$) of the training dataset is used for validation. For evaluation, we employ a total of 1,276 songs from the MAESTRO dataset\cite{hawthorne2018enabling}. As several state-of-the-art audio transcription models are trained on MAESTRO\cite{hawthorne2017onsets,kong2021high}, it can be considered a high-quality ground truth testbed for the raw MIDI data. 

\noindent\textbf{Implementation.}
We train the GPT-2 architecture containing 12 hidden layers implemented by Huggingface\cite{wolf2019huggingface}. We use the AdamW optimizer with a learning rate initially set at 1e-4, which is linearly reduced to 0 after 1,000 warm-up steps. We set the maximum token length as 1024 and run a total of 144,786 training iterations or 6 epochs. We infer all test samples using token prompts from MAESTRO, sampling each prompt five times to reduce contingency in random generations. From each song of MAESTRO, we take the beginning of a song as a prompt and examine the prompt token lengths of $2^l$ with $l\in\{4,6,8\}$ (The lengths in time are about 2, 5, 15 seconds, respectively). We use top-$k$ sampling with $k=32$ until the sample reaches 1 minute maximum. 

\noindent\textbf{Compared Methods.}
To study the effect of structural embeddings, we train and test three variants of GPT-2. \textbf{GPT2} is the vanilla language model without any structural embeddings. \textbf{GPT2-RE} is GPT2 equipped with four structural embeddings, all initialized randomly. \textbf{GPT2-SE} is the same as GPT2-RE, but with part and time embeddings initialized using sinusoidal weights instead.

\noindent\textbf{Objective Evaluation.} % specific contents to appendix
We use three metrics to evaluate the influence of the structural embeddings on both temporal and harmonic structures of the generated music. \textit{Structureness Indicator (SI)} measures the largest degree of repeatedness among various intervals in generated music. We exploit this metric to evaluate how much a generated sequence reflects the repeating nature of music according to the structural embeddings\cite{wu2020jazz}. It is computed by a fitness score based on a similarity matrix\cite{mueller2011segment, wu2020jazz}. We convert the generated samples to audio, set the sampling rate at 1 Hz, and compute the score in the unit of seconds on the repeated duration intervals in $3\sim 8$, $8\sim 15$, and $15\sim N$ seconds, where $N$ is the total length of a song \cite{wu2020jazz}. We also use two existing metrics to measure the harmonic structure of the generated sequence. \textit{Chord Progression Variation Rationality (CPVR)} measures the rationality of chord changes within the generated harmonic context\cite{zhang2022structure}. It is computed as the conditional probability of the existence of each unique chord $n$-gram given its previous $n$-gram within the generated music. We consider each chord $n$-gram unique if at least one chord element is different from other $n$-grams. \textit{Chord Progression Irregularity (CPI)} measures the ratio of unique chord $n$-grams from generated music, compensating CPVR that yields high scores with frequently occurring chords \cite{wu2020jazz}. These two chord-related metrics are examined when $n \in \{2,3,4\}$. We utilize the existing rule-based algorithm \cite{huang2020pop} to label chords every 500 milliseconds or 1 second throughout the song. We do not use beat-based intervals which cannot be determined without beat or bar annotations. For more details on implementing these metrics, please refer to Appendix~\ref{sec:metric_details}.
    
\noindent\textbf{Subjective Evaluation.}
We conduct the common A-vs-B human-rating procedure\cite{huang2018music, yu2022museformer, agostinelli2023musiclm}. We randomly select 12 prompts of length 64 from MAESTRO and generate the samples in a maximum of 30 seconds. For each prompt, samples from the four different models including the ground truth from MAESTRO are alternatively compared under the A/B test procedure. Four raters score each sample pair based on how natural each sample sounds (\textit{Naturalness}) and how well each sample preserves the characteristics of the given prompt (\textit{Prompt Maintenance}). The measurement of prompt maintenance aims to validate structural fidelity in music. Structural fidelity has been defined by how a given phrase or theme is repeated and developed throughout the composition \cite{wu2020jazz, zhang2022structure, shih2022theme}. Consequently, each rater listens to a total of 72 pairs where 6 model pairs are compared for 12 samples.

\subsection{Objective Results}

\noindent\textbf{Repeatedness of Musical Patterns.}
\tabref{tab:objective} shows that GPT2-SE outperforms other methods in Structureness Indicators (SI) for all interval lengths and prompt lengths. It demonstrates that sinusoidal initialization may reliably generate repeated patterns in various lengths, regardless of prompt length. GPT2-RE shows better performance than GPT2 with prompt lengths longer than 16. It is unstable compared to other models showing large degradation with prompts at length 16. It indicates that GPT2-RE may be more beneficial than GPT2 in generating repeated music only from the prompts longer than 5 seconds. In contrast, removing supplementary embeddings may be better than preserving them for creating regular patterns, only from prompt lengths shorter than 5 seconds.

\noindent\textbf{Rationality and Irregularity of Chord Progression.}
\tabref{tab:objective} shows that GPT2-SE receives the highest scores in Chord Progression Variation Rationality (CPVR) among all settings, but scores the lowest in Chord Progression Irregularity (CPI). It indicates that sinusoidal initialization can induce more common chords, which can lead to low diversity of the overall harmonic context compared to the other models. This compromise is further substantiated by the data in \figref{fig:corr}, which shows a robust inverse correlation between CPI and CPVR. Moreover, high SI values may be associated with high CPVR when certain harmonic contexts are repeated. GPT2-RE shows the worst CPVR scores in prompt lengths 16, while CPI scores are shown the best (\figref{fig:corr}). For longer prompts, GPT2-RE gets better CPVR scores than GPT2 while showing worse CPI scores. GPT2 reveals the best CPI scores with prompt lengths 64 and 256. It demonstrates that randomly initialized structural embeddings may benefit in making rational chords over the vanilla GPT-2 with prompts longer than 5 seconds while creating more unique chords with shorter prompts.

\begin{table}
    \centering
    \resizebox{\columnwidth}{!}{
        \begin{tabular}{l|cccc}
            \toprule
            PNSR($s$) & $s<0.25$ & $0.25\leq s<0.5$ & $0.5\leq s<0.75$ & $0.75\leq s$\\ 
            \midrule
            Prompt-64 & 59 & 191 & \textbf{433} & \textbf{593}\\
            Prompt-16 & \textbf{329} & \textbf{397} & 274 & 276 \\ 
            \bottomrule
        \end{tabular}
    }
    \caption{Number of prompts in different PNSR ranges.}
    \label{tab:pnsr}
\end{table}

\begin{figure}
\centering
\includegraphics[width=\columnwidth]{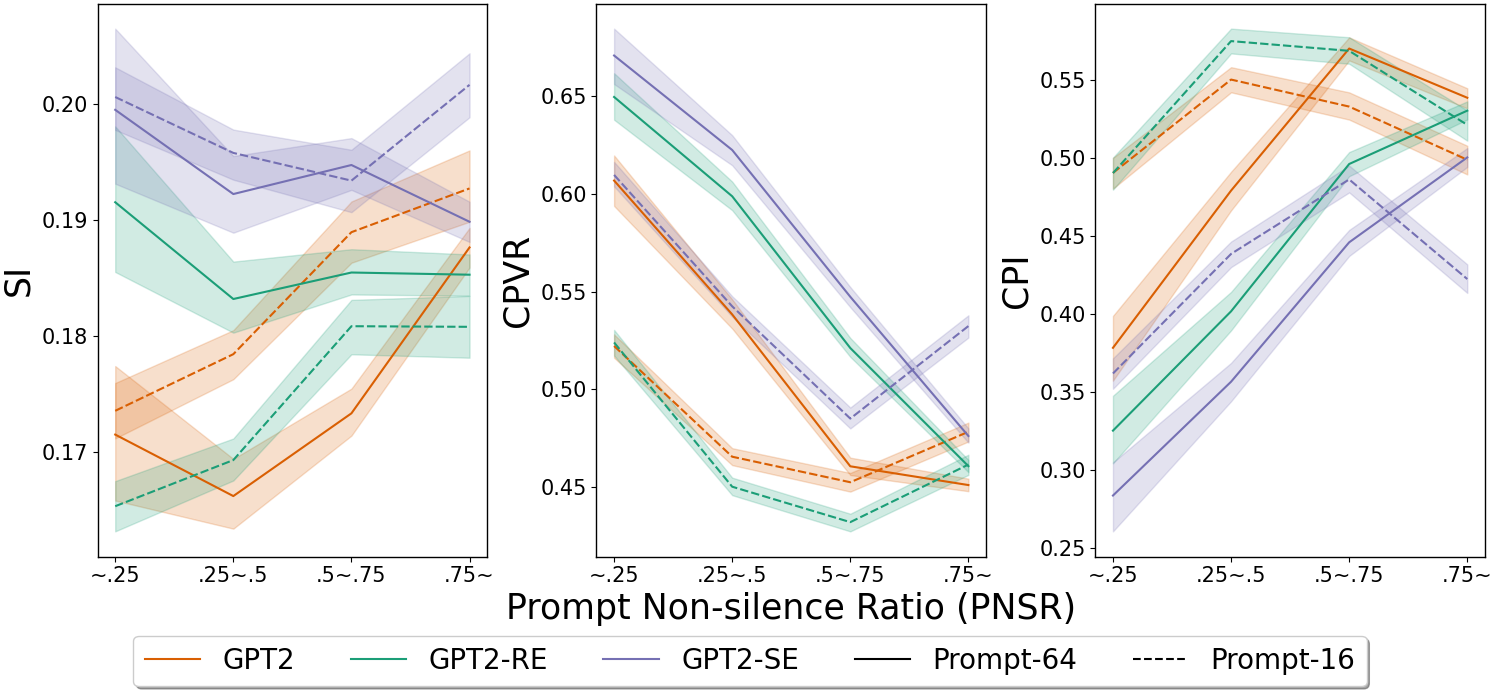}
\caption{PNSR vs. SI, CPVR, and CPI plots with prompt lengths 64 (solid) and 16 (dashed) from the three methods.}
\label{fig:pnsr}
\vspace{-.1in}
\end{figure}

\noindent\textbf{Periods of Silence in Prompts.} 
We closely investigate the reason for the large degradation in performance of GPT2-RE at the prompt length 16. We hypothesize that prompts of this length become distinguished in some aspect from the longer prompts, and GPT2-RE is particularly sensitive to this corresponding aspect. To verify this assumption, we compute the ratio of non-silence periods within the prompts (shortened as PNSR) with lengths 16 and 64. We convert each prompt to a $P \times T$ 2-dimensional piano roll, where $P$ is the number of MIDI pitches, and $T$ is the number of time frames. Then, we obtain PNSR as the number of non-zero columns divided by the length of the piano roll. \tabref{tab:pnsr} shows the distribution of the prompts across 4 ranges of the PNSR. The number of prompts with low PNSR under 0.5 increases when its length is 16, compared to 64. \figref{fig:pnsr} demonstrates relationships between PSNR and the three metric scores of all models averaged across the sub-metrics. All metric scores of GPT2-RE abruptly decline as the PSNR decreases under 0.5 when the prompt length is 16, in contrast to other models revealing relatively minor changes across the prompt length. It signifies that GPT2-RE may be more responsive than the other models to the amount of silence in the input prompt, which depends on the prompt lengths.

\begin{figure*}
\begin{subfigure}{.25\linewidth}
  \centering
  \includegraphics[width=.95\linewidth]{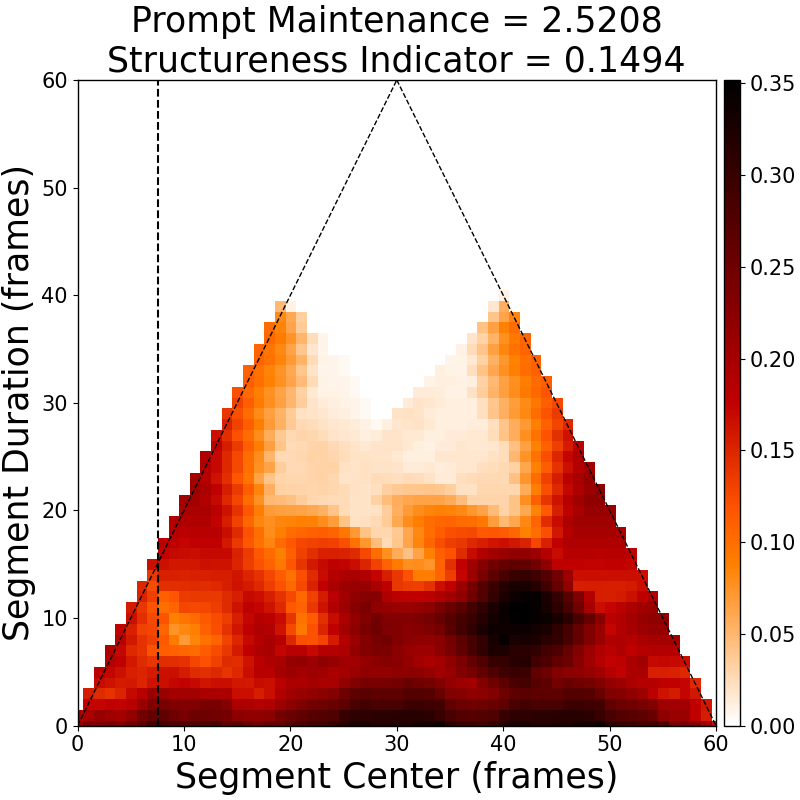}
  \caption{GPT2}
  \label{fig:sp_gpt2}
\end{subfigure}%
\begin{subfigure}{.25\linewidth}
  \centering
  \includegraphics[width=.95\linewidth]{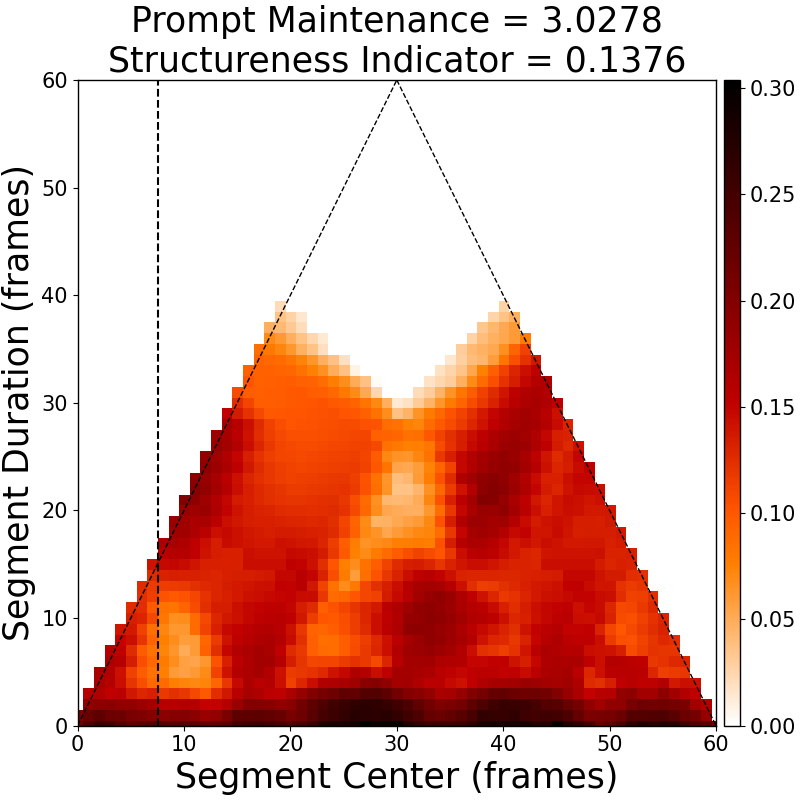}
  \caption{GPT2-RE}
  \label{fig:sp_gpt2-re}
\end{subfigure}%
\begin{subfigure}{.25\linewidth}
  \centering
  \includegraphics[width=.95\linewidth]{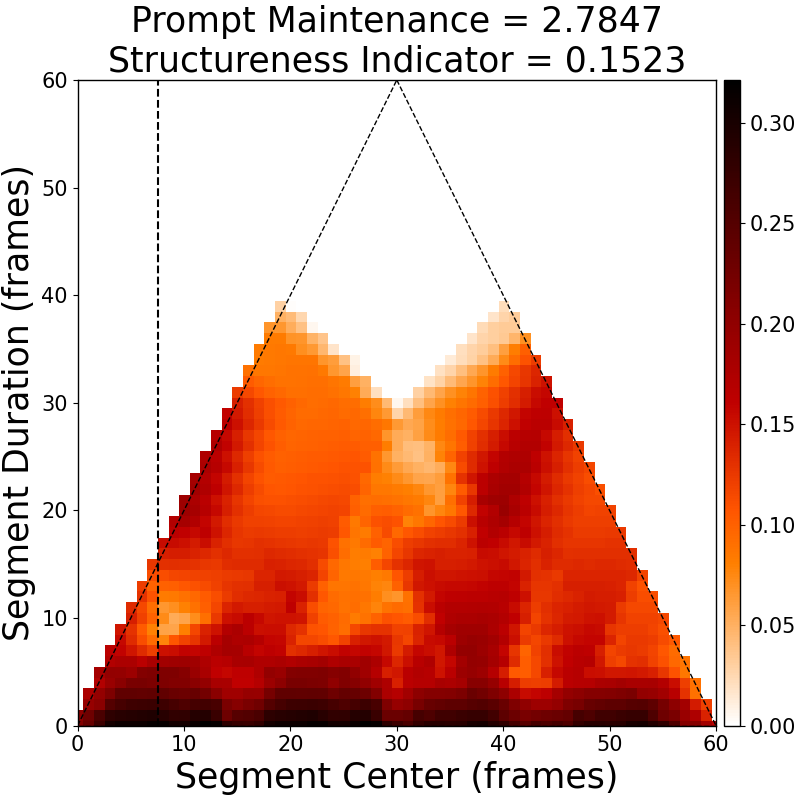}
  \caption{GPT2-SE}
  \label{fig:sp_gpt2_se}
\end{subfigure}%
\begin{subfigure}{.25\linewidth}
  \centering
  \includegraphics[width=.95\linewidth]{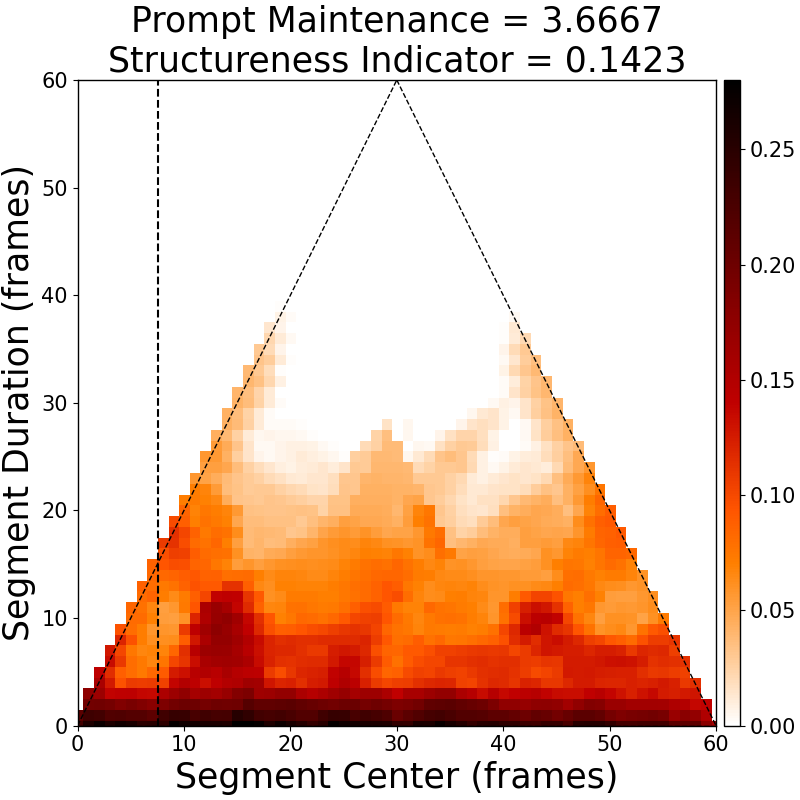}
  \caption{GT}
  \label{fig:sp_gt}
\end{subfigure}
\caption{Compiled fitness scape plots across 12 samples from the four models: Each plot displays the maximum fitness scores among all stacked scores from 12 samples for each entry within the plot. Each plot illustrates the segment center in frames on the X-axis and the segment duration in frames on the Y-axis. Every black dotted line marks the average end time (3.77 seconds) across the prompts. Additionally, average scores for Structureness Indicators and Prompt Maintenance across the 12 samples are provided at the top of each plot.}
\label{fig:fitness}
\vspace{-.1in}
\end{figure*}

\begin{table}
    \centering
    \resizebox{\columnwidth}{!}{
        \begin{tabular}{l|ccc|c}
            \toprule
            Method & GPT2 & GPT2-RE & GPT2-SE & GT\\ 
            \midrule
            Naturalness & 30 & \textbf{63} & 37 & 109\\
            Prompt Maintenance & 31 & \textbf{57} & 46 & 92\\ 
            \bottomrule
        \end{tabular}
    }
    \caption{Win counts from the subjective evaluation.}
    \label{tab:subjective}
\end{table}

\begin{table}
    \centering
    \resizebox{\columnwidth}{!}{
        \begin{tabular}{c|ccc|c}
            \toprule
            \multicolumn{1}{l|}{Metric} & \multicolumn{4}{c}{Naturalness}\\
            \midrule
            \multicolumn{1}{l|}{Method Pair} & Wins & Ties & Losses & $p$-value\\ 
            \midrule
            \multicolumn{1}{l|}{GPT2-RE vs. GPT2-SE} & \textbf{24} &	14	&10	&0.0033 \\
            \multicolumn{1}{l|}{GPT2-RE vs. GPT2} & \textbf{31}	&10	&7&	0.0004\\ 
            \multicolumn{1}{l|}{GPT2-RE vs. GT} & 8	&5&	\textbf{35}	&0.0011\\ 
            \multicolumn{1}{l|}{GPT2-SE vs. GPT2} & 21	&8	&19	&0.3692\\ 
            \multicolumn{1}{l|}{GPT2-SE vs. GT} & 6	&7	&\textbf{35}&	5.11e-6\\ 
            \multicolumn{1}{l|}{GPT2 vs. GT} & 4	&5&	\textbf{39}&	1.93e-6\\ 
            \toprule
            \multicolumn{1}{l|}{Metric} & \multicolumn{4}{c}{Prompt Maintenance} \\
            \midrule
            \multicolumn{1}{l|}{Method Pair} & Wins & Ties & Losses & $p$-value\\ 
            \midrule
            \multicolumn{1}{l|}{GPT2-RE vs. GPT2-SE} & 22	&11	&15	&0.1939 \\
            \multicolumn{1}{l|}{GPT2-RE vs. GPT2} & \textbf{24}	&15	&9	&0.0115\\ 
            \multicolumn{1}{l|}{GPT2-RE vs. GT} & 11	&8	&\textbf{29}&	0.0025\\ 
            \multicolumn{1}{l|}{GPT2-SE vs. GPT2} & 19	&13	&16	&0.5246\\ 
            \multicolumn{1}{l|}{GPT2-SE vs. GT} & 12	&6	&\textbf{30}&	0.0064\\ 
            \multicolumn{1}{l|}{GPT2 vs. GT} & 6&	9	&\textbf{33}&	1.38e-5\\ 
            \bottomrule
        \end{tabular}
    }
    \caption{Pairwise comparison results of the subjective evaluation scores on Naturalness and Prompt Maintenance.}
    \label{tab:subjective_pairwise}
\vspace{-.1in}
\end{table}

\subsection{Subjective Results}\label{sec:subjective}

\tabref{tab:subjective} indicates that GPT2-RE achieves the highest number of wins among the models, while GPT2 records the lowest. \tabref{tab:subjective_pairwise} presents pairwise comparisons among the models regarding Naturalness and Prompt Maintenance, including $p$-values from the Wilcoxon signed-rank test based on the scores collected by the raters \cite{huang2018music, yu2022museformer}. In Naturalness, all model pairs except for GPT2-SE and GPT2 demonstrate significant differences ($p<.005$). Concretely, GPT2-RE significantly beats GPT2, whereas GPT2-SE does not show a significant difference from GPT2. In Prompt Maintenance, GPT2-RE also significantly outperforms GPT2 ($p<.05$), while GPT2-SE does not. These results suggest that incorporating randomly initialized structural embeddings significantly improves the ability of the basic GPT-2 to create plausible and prompt-consistent music, whereas sinusoidal initialization shows no meaningful difference.

These results can give us insight that the recurrence of specific passages, which can measured by SI, may not simply exhibit a correlation with the sense of prompt maintenance. Alternatively, scores for Prompt Maintenance are demonstrated to be associated with those of Naturalness, as evidenced by an examination of the Pearson's correlation coefficient between the scores of these two metrics across the models ($r = 0.6559$). Moreover, it may imply that the least trade-off between CPVR and CPI scores shown in GPT-RE can be beneficial in reaching higher naturalness and prompt maintenance compared to the other two models revealing extreme trade-offs. In the following section, we investigate how GPT2-RE can be superior in listeners' evaluation to the other models in terms of prompt maintenance. 

\subsection{Fitness Scape Plots}\label{sec:analysis}

We conduct a detailed investigation of the fitness scape plot that spans the entire song \cite{mueller2012scape}. The fitness scape plots are derived from all 12 samples of each model. We set each frame at 500 ms to examine the plots at a higher resolution. Then, we compile the plots to calculate the maximum fitness scores from all stacked scores for each entry within the plot. The X-axis and Y-axis of the plot denote the center and the length of the repeated segments within the song, termed 'segment center' and 'segment duration,' respectively. We use the maximum scores rather than the average to clearly observe the model’s peak performance in creating structural patterns across various segment durations.  

\figref{fig:fitness} displays the resulting fitness scape plots of the three models and GT. The fitness scape plot for GT is distinctive compared to those of the other models. It shows clearly defined regular patterns, particularly darker at the side edges of the triangle. Furthermore, the scores are relatively low for short segment durations compared to those of other models. This suggests that GT may feature musical patterns in longer durations, developed from the given prompts, leading to the highest Prompt Maintenance (PM) scores. While GPT2-RE and GPT2-SE seem similar, GPT2-RE reveals higher fitness scores at the upper side of the plot than GPT2-SE and lower fitness scores under the plot. GPT2-RE also has lower fitness scores for the shorter segments close to the prompt (at the lower left corner), similar to GT. Conversely, GPT2-SE demonstrates fitness scores that are broadly distributed below the plot. It indicates that GPT2-RE may be more beneficial than GPT2-SE to generate longer patterns derived from the prompt leading to higher PM scores. GPT2 shows a score pattern that completely neglects the given prompt: the fitness scores are concentrated in the last half of the song, yielding highly irregular patterns compared to other models. This aspect clarifies why GPT2 obtains the worst PM score. We also note that the aligned Structureness Indicators (SI) scores are not highly correlated to PM scores  ($r=-0.1429$). GPT2 and GPT2-SE show higher average scores for SI than GT, unlike for PM. It may imply that repetition of the musical patterns cannot solely represent how well the given prompt is retained in the music. This leads to an intuition suggesting the necessity for additional structural metrics that can grasp the high-level fidelity in the musical structure, beyond the simple repetition of musical patterns.

\section{Conclusion}\label{sec:conclusion}
We have reconstructed practical structural embeddings for raw MIDI data, which lacks domain-specific annotations. We have verified that adding structural embeddings enhances the model's performance in capturing the structural aspects of music compared to the model without them. It is encouraging that structural embeddings can be easily integrated with existing language models such as GPT-2. Under various embedding setups, we find that: 1) random initialization generates more plausible music for human listeners but lacks stability; 2) sinusoidal initialization of temporal embeddings helps produce repeated patterns and common chords in music. These distinct strengths of the two settings may enable users to enhance specific aspects of music in certain ranges when creating their products. Our demonstration\footnote{\url{https://free-pig-6c6.notion.site/ISMIR2024-Demo-Page-7d640a34529646ee91acc6c41ef1ec27?pvs=4}} showcases the corresponding examples. 

% For bibtex users:
\bibliography{main}

\appendix\label{sec:appendix}
\section{Related Work}\label{sec:related_work}

% For interested readers, here we discuss a list of related works on symbolic music generation approaches based on the Transformer architecture and various methods used to capture structural information in musical data or sequences in general.

\subsection{Transformer-based Symbolic Music Generation.} Inspired by previous work on large-scale text generation, there exist multiple frameworks that leverage the Transformer architecture for event-based music generation. Among many, Music Transformer\cite{huang2018music} was the first to successfully apply Transformers to music generation with long-range dependencies using relative local attention. In the following year, MuseNet\cite{payne2019musenet} proposed applying the GPT2\cite{radford2019language} architecture to MIDI-based music generation, showing state-of-the-art performance under prompts with various contexts such as composer and instrumentation. Several works have also studied music generation in specific genres such as Pop Music Transformer\cite{huang2020pop} and Jazz Transformer\cite{wu2020jazz}. MusicBERT\cite{zeng2021musicbert} proposed using bidirectional masked token prediction as in BERT\cite{devlin2018bert} with bar-level masking to avoid information leakage among nearby tokens. Museformer\cite{yu2022museformer} proposed using a coarse- and fine-grained attention module to reduce the quadratic computational cost of attention to nearly linear. In our work, we implement a revamped version of MuseNet\cite{payne2019musenet} to cope with the lack of reproducible code, and also empirically study different design choices on structural embedding initializations and prompt lengths during inference.

\subsection{Structural Embeddings for Music Generation.} For sequence modeling, the original Transformer architecture proposed an absolute sinusoidal positional encoding\cite{vaswani2017attention}. To cope with long sequences, T5\cite{raffel2020exploring} and Transformer-XL\cite{dai2019transformer} have proposed a relative positional encoding that adds inductive bias directly to the attention-score matrix. Extending to vision- and graph-data, ViT\cite{dosovitskiy2020image} and Graph Transformer\cite{dwivedi2020generalization} also proposed positional encoding modules that leverage only topological data structure, meaning that it can be applied to any large-scale datasets without additional knowledge. In the musical domain, there exist many variations of event-based MIDI tokenization pipelines such as REMI\cite{huang2020pop}, CP\cite{hsiao2021compound}, and OctupleMIDI\cite{zeng2021musicbert} that aggregate structural MIDI information into a single token. While they effectively reduce the computational cost by shortening the overall sequence length, these require additional structural information that is often missing in in-the-wild music data such as chord and bar structure. Therefore, we focus on the structural embedding procedure for event-based MIDI tokenization that does not require any human annotation or heuristic labeling.

\section{Metric Details}\label{sec:metric_details}
\subsection{Structureness Indicator.} We leverage Libfmp API\footnote{\url{https://meinardmueller.github.io/libfmp/build/html/index_c4.html}} to compute SI and the corresponding fitness scape plots. We calculate SI as the overall fitness score following the procedure provided by the authors\footnote{\url{https://www.audiolabs-erlangen.de/resources/MIR/FMP/C4/C4S3_ScapePlot.html}} and parameter settings from\cite{wu2020jazz}: we use the sampling rate of 1 Hz, or 1 second per frame, for computing the objective metric scores. For the fitness scape plot, we compute a $N\times N$ fitness matrix from each of the 12 listening samples for each model, where $N$ is the maximum number of frames derived from downsampling the original audio files, and each frame denotes 500 ms for fine-level visualization\cite{mueller2014si}. Then, we average the fitness matrices by sample. Empirically, we set $N=61$ as the maximum length of the samples in 30 seconds. If a sample has a length smaller than $N$, we pad the remaining length with NaN values.

\subsection{Chord Progression Variation Rationality.} We follow\cite{zhang2022structure} to compute this metric. The conditional probability $V_i$ is approximated from the appearance frequency of a unique chord $n$-grams for each $n\in\{2,3,4\}$ within the training set, which is the aggregation of Pop1k7\cite{hsiao2021compound} and GiantMIDI-Piano\cite{kong2020giantmidi}, as follows:
\begin{equation}
    V_i=\mathbf{p}\left(C_i^{i+n} \mid C_i^{i-1+n}\right)
\end{equation}
where $i$ is the chord index, $C_i^{i+n}$ is a current chord $n$-gram spanning from the $i$-th chord to the $i+n$-th chord, and $\mathbf{p}\left(\cdot \mid \cdot\right)$ is the conditional probability.

\end{document}